\documentclass[runningheads]{llncs}
\usepackage{graphicx}
\usepackage{booktabs}
\usepackage{caption}
\usepackage{rotating}
\usepackage{multirow}
\usepackage{booktabs}
\usepackage{adjustbox}
\usepackage{pifont}
\usepackage{amsmath}
\usepackage{colortbl} 
\definecolor{lightyellow}{RGB}{255, 255, 204} 
\definecolor{lightgreen}{RGB}{204, 255, 204} 
\definecolor{lightgray}{RGB}{220, 220, 220}    
\usepackage{xcolor}   
\usepackage{graphicx} 
\usepackage{multirow} 
\usepackage{makecell} 
\usepackage{marvosym}

\usepackage{graphicx} 
\usepackage{transparent} 
\begin{document}
\vspace{-5cm}   
\title{Self is the Best Learner: CT-free Ultra-Low-Dose PET Organ Segmentation via Collaborating Denoising and Segmentation Learning}
\titlerunning{LDOS}
\author{
    Zanting Ye\textsuperscript{1,*},
    Xiaolong Niu\textsuperscript{1,*},
    Xu Han\textsuperscript{2,*},
    Xuanbin Wu\textsuperscript{1},
    Wantong Lu\textsuperscript{1},
    Yijun Lu\textsuperscript{1},
    Hao Sun\textsuperscript{1},
    Yanchao Huang\textsuperscript{3},
    Hubing Wu\textsuperscript{3},
    Lijun Lu\textsuperscript{1,4,5,6,\Letter}
}

\authorrunning{Ye et al.}
\institute{
    \textsuperscript{1}School of Biomedical Engineering, Southern Medical University, Guangzhou, China \\
    \textsuperscript{2}School of Biomedical Engineering, Shanghai Jiao Tong University, Shanghai, China \\
    \textsuperscript{3}Nanfang PET Center, Nanfang Hospital Southern Medical University, Guangzhou, China \\
    \textsuperscript{4}Guangdong Provincial Key Laboratory of Medical Image Processing, Southern Medical University, Guangzhou, China \\
    \textsuperscript{5}Guangdong Province Engineering Laboratory for Medical Imaging and Diagnostic Technology, Southern Medical University, Guangzhou, China \\
    \textsuperscript{6}Pazhou Lab, Guangzhou, China \\
    \email{ljlubme@gmail.com}\\
}
\renewcommand{\thefootnote}{\fnsymbol{footnote}}
\footnotetext[1]{These authors contributed equally to this work.}


%
%
\maketitle        
\begin{abstract}
Organ segmentation in Positron Emission Tomography (PET) plays a vital role in cancer quantification. Low-dose PET (LDPET) provides a safer alternative by reducing radiation exposure. However, the inherent noise and blurred boundaries make organ segmentation more challenging. Additionally, existing PET organ segmentation methods rely on co-registered Computed Tomography (CT) annotations, overlooking the problem of modality mismatch. In this study, we propose LDOS, a novel CT-free ultra-LDPET organ segmentation pipeline. Inspired by Masked Autoencoders (MAE), \textit{we reinterpret LDPET as a naturally masked version of Full-Dose PET (FDPET)}. LDOS adopts a simple yet effective architecture: a shared encoder extracts generalized features, while task-specific decoders independently refine outputs for denoising and segmentation. By integrating CT-derived organ annotations into the denoising process, LDOS improves anatomical boundary recognition and alleviates the PET/CT misalignments. Experiments demonstrate that LDOS achieves state-of-the-art performance with mean Dice scores of 73.11\% (\textsuperscript{18}F-FDG) and 73.97\% (\textsuperscript{68}Ga-FAPI) across 18 organs in 5\% dose PET. Our code will be available at https://github.com/yezanting/LDOS.

\end{abstract}
\section{Introduction}
Positron emission tomography (PET) is a powerful molecular imaging modality that visualizes radiotracer distribution to reveal physiological processes. Low-dose PET (LDPET) reduces radiation exposure and has demonstrated diagnostic equivalence to Full-Dose PET (FDPET)~\cite{zhou2021mdpet,zhou2020supervised,liu2022short,chen2022evaluation}. However, its clinical adoption remains limited by the lack of robust quantitative tools in high-noise settings. Accurate tracer uptake measurements in non-tumoral organs are critical for tumor quantification within Volumes of Interest (VOIs) \cite{aoki2001fdg,mansouri2024development}. LDPET organ segmentation offers a promising approach for tracer uptake assessment and kinetic measurements \cite{zaidi2017towards,ziai2016role}, yet it remains relatively understudied.

\begin{figure*}[t]
\centering
\includegraphics[width=0.75\textwidth]{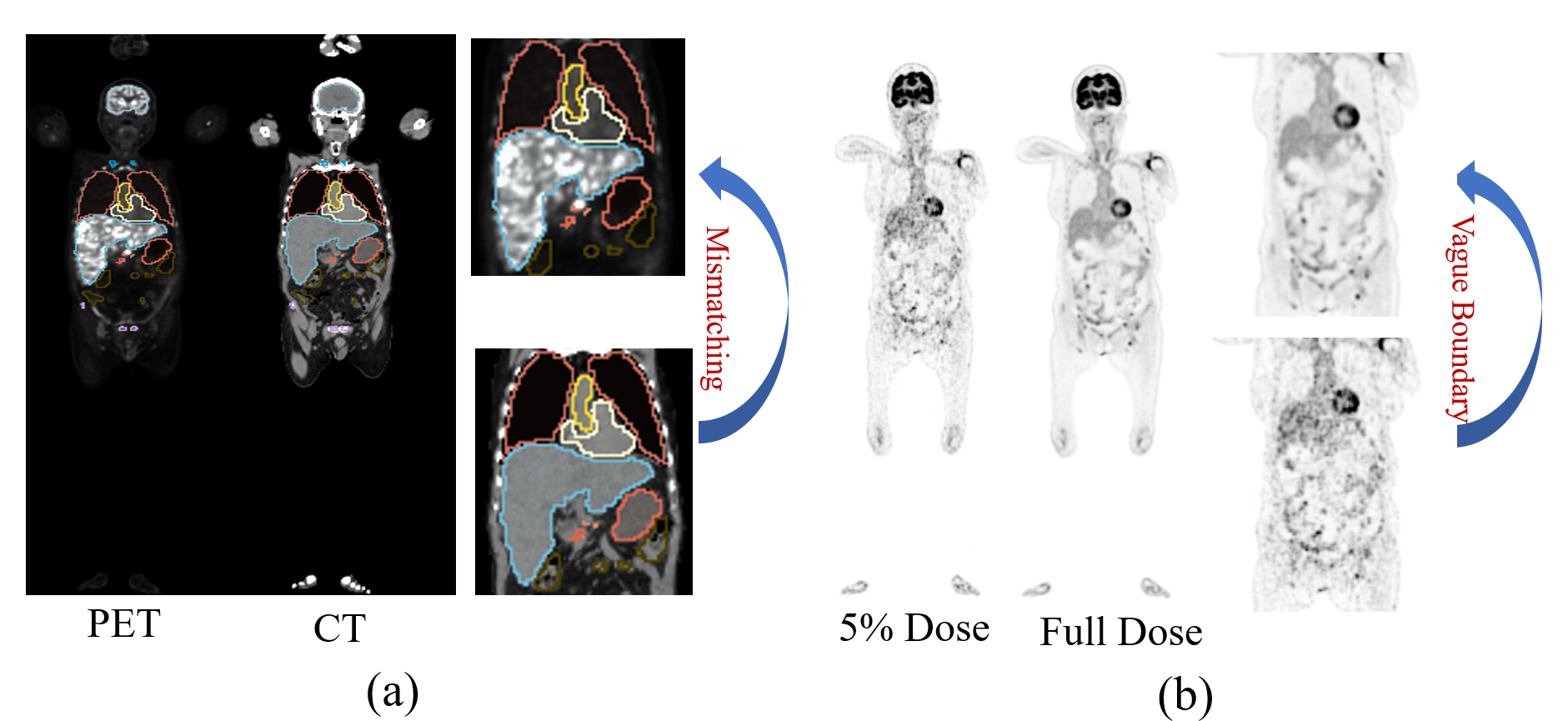} 
\caption{(a) Misalignments between PET and CT caused by respiratory motion (e.g., liver displacement). (b) Noise and blurred boundary artifacts in LDPET.}
\label{fig1}
\vspace{-0.5cm}  
\end{figure*}

In this study, we summarize the challenges of LDPET organ segmentation: (1) \textbf{Data scarcity}: Clinical FDPET dominance limits LDPET dataset availability. (2) \textbf{Annotation complexity}: Low soft-tissue contrast in PET, worsened in LDPET, makes organ annotation challenging. Existing PET organ segmentation methods rely on co-registered Computed Tomography (CT) annotations, overlooking the issue of modality mismatch (Fig. \ref{fig1}a). (3) \textbf{Noise-induced ambiguity}: Blurred anatomical boundaries hinder accurate segmentation (Fig. \ref{fig1}b).

To address these challenges, we propose LDOS, a novel CT-free LDPET organ segmentation pipeline via collaborating denoising and segmentation learning. Inspired by MAE \cite{he2022masked}, we reinterpret LDPET as a naturally masked version of FDPET, where the denoising process inherently recovers organ-level semantics. Recent studies have demonstrated that incorporating organ annotations into LDPET denoising can improve FDPET reconstruction, highlighting the shared semantic priors between denoising and segmentation tasks \cite{jiang2023semi,zhang2023hierarchical,dayarathna2024deep}.

Unlike traditional MAE-based methods that depend on large-scale pretraining, LDOS employs self-denoising to reinforce organ-level semantic learning, reducing the risks of overfitting and bias-learning in small-scale PET datasets~\cite{wald2024revisiting,munk2024amaes,rokuss2024fdg}. From a technical perspective, LDOS uses a simple yet effective training pipeline. A shared encoder extracts generalized features for both tasks, while task-specific decoders independently refine outputs for denoising (supervised by FDPET) and segmentation (supervised by CT annotations during training only). By incorporating CT-derived organ annotations into the denoising process, LDOS improves anatomical boundary recognition and mitigates PET/CT misalignments.

\begin{figure*}[t]
\centering
\includegraphics[width=1\textwidth]{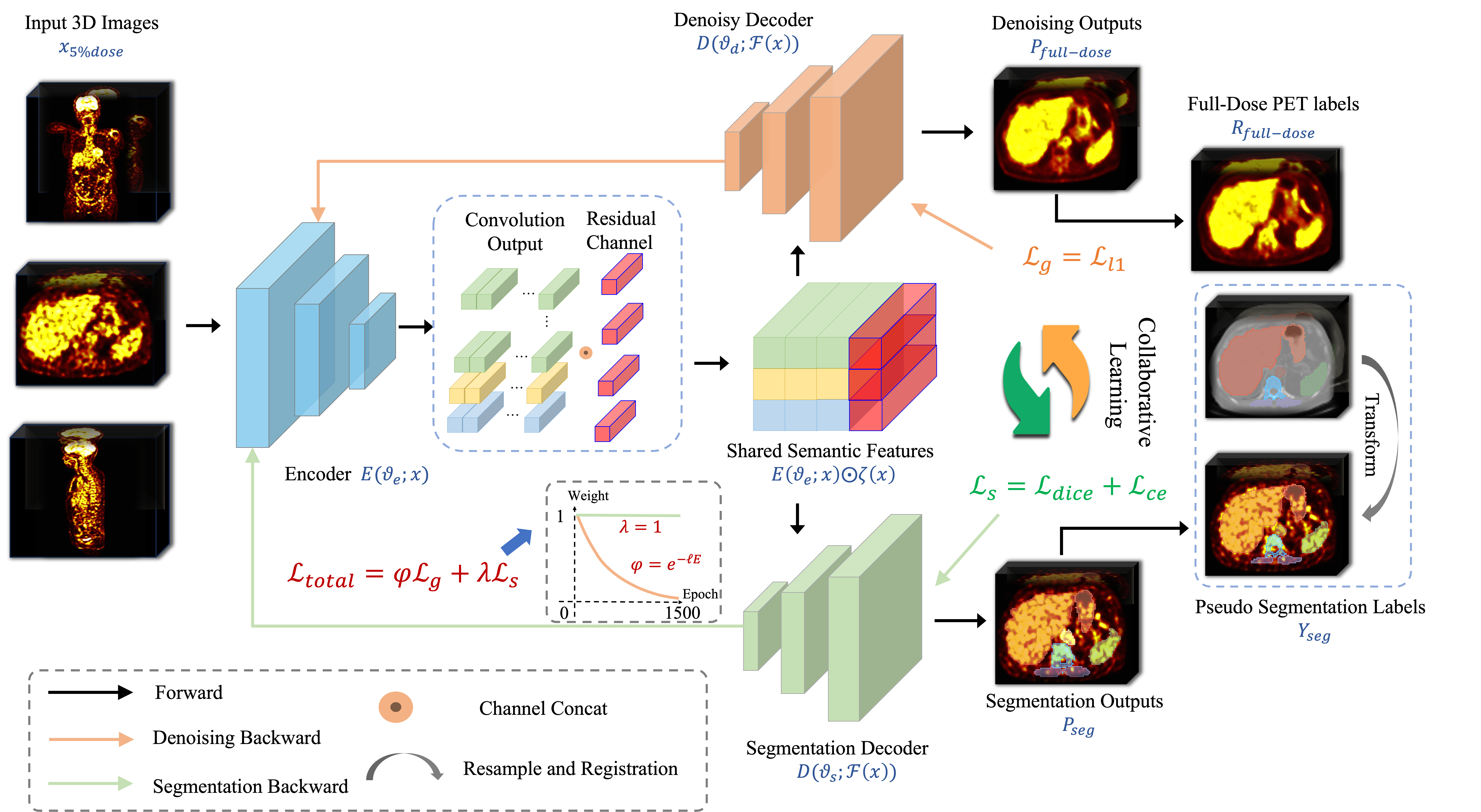} 
\caption{Overview of LDOS. LDOS employs the shared encoder and independent decoders for denoising and segmentation tasks.}
\label{fig2}
\vspace{-0.5cm}  
\end{figure*}

\section{Related Work}
\subsubsection{PET Organ Segmentation}
Existing PET segmentation methods primarily focus on organs with high tracer uptake (e.g., tumors) in FDPET, while organ segmentation in LDPET remains largely unexplored \cite{taghanaki2019combo,oh2020semantic,bao2024ct}. Furthermore, PET organ segmentation methods rely on co-registered CT annotations, ignoring modality mismatch \cite{RN14,wang2025robust}. To address this gap, we propose LDOS, which operates directly on noisy ultra-LDPET data. By incorporating CT-derived organ annotations into the denoising process, LDOS improve anatomical boundary recognition and mitigates PET/CT misalignments.

\subsubsection{MAE in Medical Image Analysis}
MAE excels in natural images with large-scale pretraining (e.g., ImageNet \cite{deng2009imagenet}). However, medical imaging lacks comparable datasets, making pretraining on limited data prone to overfitting and bias \cite{wald2024revisiting,munk2024amaes,rokuss2024fdg}. Inspired by MAE, where masking removes random patches, we reinterpret LDPET as positional masking of FDPET, where low-dose acquisition inherently “masks” high-frequency details. Instead of resource-intensive pretraining, LDOS integrates denoising and segmentation into a single-stage collaborative learning process, avoiding small-data pitfalls and ensuring anatomical consistency.

\section{Main Methodology}
LDOS employs a simple yet effective training pipeline, which improves segmentation accuracy by jointly reconstructing FDPET signals. As shown in Fig.~\ref{fig2}, it employs a shared encoder $E_\theta$ and two task-specific decoders $D_\phi$  and $D_\psi$.

\subsection{Model Architecture}
\subsubsection{Shared Encoder}
LDOS uses the nnU-Net backbone~\cite{isensee2021nnu1} with the ResEncL configuration~\cite{isensee2024nnu}. Inspired by MAE, LDOS employs the LDPET denoising process to learn semantic latent representations. MAE uses the patch-level discretized images as inputs~\cite{dosovitskiy2020image}. In the medical imaging context, large-scale patch masking leads to excessive information loss. The degradation in LDPET physically stems from lower photon counts and consequent statistical noise. LDOS approximates this effect as a pixel-level masking process, which facilitates the retention of core information:
\begin{equation}
{{x}_{i}} = \psi (\Lambda {{r}_{i}}, k)
\end{equation}
where \({{x}_{i}}\) is the LDPET image, \({{r}_{i}}\) is the FDPET image, and \(\Lambda r\) represents the pixel-level discretization. The encoder processes the input \({{x}_{i}}\) to extract semantic latent features:
\begin{equation}
F({{x}_{i}}) = E({{\vartheta }_{e}};{{x}_{i}})\odot \zeta ({{x}_{i}})
\end{equation}
where \(F({{x}_{i}})\) represents shared semantic features, \(E\) is the encoder, \({{\vartheta }_{e}}\) denotes encoder parameters, \(\odot\) signifies dimensionality stacking, and \(\zeta\) is the residual connection. Unlike standard MAE approaches that focus on visible pixels, our method models all pixels, providing comprehensive semantic representation.

\subsubsection{Segmentation and Denoising Decoders}
The encoder output feeds into both segmentation and denoising decoders, which share a consistent architecture. To enhance feature learning, a deep supervision strategy is incorporated ~\cite{isensee2021nnu1}:
\begin{equation}
P_{full-dose}^{z} = D_d(\vartheta _{d}^{z};F({{x}_{i}}))
\end{equation}
\begin{equation}
P_{seg}^{z} = \frac{\exp ({{D}_{s}^{c}}(\vartheta _{s}^{z};F({{x}_{i}})))}{\sum\limits_{C}{\exp ({{D}_{s}^{c}}(\vartheta _{s}^{z};F({{x}_{i}}))}}
\end{equation}
where \(P_{full-dose}^{z}\) is the FDPET prediction at scale \(z\), \(D\) is the decoder, and \(\vartheta _{d}^{z}\) denotes the decoder parameters. Similarly, \(P_{seg}^{z}\) refers to the segmentation prediction for the \(c\)-th class, with \({{D}_{s}^{c}}\) being the decoder for class \(c\) and \(C\) the total number of segmentation classes.

\subsection{Training via Collaborative Denoising and Segmentation}
To address PET/CT misalignments~\cite{kovacs2023addressing,bao2024ct}, misalignment data augmentation~\cite{RN14} is applied during training. This strategy mitigates label inconsistencies and improves coordination between segmentation and denoising. The denoising decoder is supervised by the loss function \({{L}_{g}}\):
\begin{equation}
{{L}_{g}} = \sum\limits_{Z}{{{\omega }^{z}} \left| P_{full-dose}^{z} - r \right|}
\end{equation}
where \(Z\) is the total number of scales, \({{\omega }^{z}}\) is the weight of scale \(z\), and \(r\) is the FDPET image. For segmentation, a combination of Dice and cross-entropy losses is used:
\begin{equation}
{{L}_{ce}} = \sum\limits_{Z}{\sum\limits_{C}{{{\upsilon }^{z}} \left(-{{y}_{c}} \log {{P}_{seg,c}^{z}} - (1 - {{y}_{c}}) \log (1 - {{P}_{seg,c}^{z}})\right)}}
\end{equation}
\begin{equation}
{{L}_{dice}} = \sum\limits_{Z}{{{\upsilon }^{z}} \left(1 - \frac{2\sum\limits_{C}{({{y}_{c}} \cdot \rho (soft\max ({{P}_{seg,c}^{z}})))}}{\sum\limits_{C}{{{y}_{c}} + \rho (soft\max ({{P}_{seg,c}^{z}}))}}\right)}
\end{equation}
\begin{equation}
{{L}_{s}} = {{L}_{ce}} + {{L}_{dice}}
\end{equation}
where \({{L}_{ce}}\) is the cross-entropy loss, \(C\) is the total number of classes, \({{\upsilon }^{z}}\) is the weight at scale \(z\), \({{y}_{c}}\) is the ground truth for class \(c\), \(\rho\) is one-hot embedding, and \({{P}_{seg,c}^{z}}\) is the segmentation prediction. The total loss function combines denoising and segmentation losses:
\begin{equation}
{{L}_{total}} = \varphi {{L}_{g}} + \lambda {{L}_{s}}
\end{equation}
where \(\varphi = e^{-\ell E}\); \(\ell = 0.002\), \(\lambda = 1\), and \(E\) denotes the number of epochs. Initially, denoising is emphasized to capture semantic features, and as training progresses, the focus shifts to refine segmentation accuracy.

\begin{table}[h]
    \centering
    \caption{Performance of the 18 organs segmentation results on the \textsuperscript{18}F-FDG and \textsuperscript{68}Ga-FAPI datasets (5-fold cross validation). Dice, IoU, HD95, and ASD are used to evaluate LDOS, and ‘±’ represents ‘mean ± standard deviation’.The green and yellow highlights mean the highest and lowest segmentation results.}
    \label{tab:merged_results}
    \setlength{\tabcolsep}{6pt} 
    \renewcommand{\arraystretch}{1.2} 
    \resizebox{\textwidth}{!}{ 
    \begin{tabular}{@{}llllllllll@{}}
        \toprule
        \multirow{2}{*}{Organs} & \multicolumn{4}{l}{\textsuperscript{18}F-FDG Dataset} & \multicolumn{4}{l}{\textsuperscript{68}Ga-FAPI Dataset} \\ 
        \cmidrule(lr){2-5} \cmidrule(lr){6-9}
        & Dice (\%) & IoU (\%) & HD95 (mm) & ASD (mm) 
        & Dice (\%) & IoU (\%) & HD95 (mm) & ASD (mm) \\ 
        \midrule
        Spleen & 76.28 ± 2.31 & 58.99 ± 3.50 & 13.56 ± 4.32 & 3.77 ± 1.16 & 69.37 ± 5.61 & 54.90 ± 5.85 & 17.67 ± 4.32 & 4.95 ± 0.88 \\
        Colon & 62.22 ± 3.47 & 53.33 ± 4.94 & 32.35 ± 8.07 & 9.42 ± 2.54 & \cellcolor{lightyellow}44.76 ± 4.79 & \cellcolor{lightyellow}29.68 ± 1.61 & \cellcolor{lightyellow}42.99 ± 7.33 & \cellcolor{lightyellow}8.58 ± 2.14 \\ 
        Urinary Bladder & 72.75 ± 7.64 & 58.61 ± 11.99 & 9.23 ± 5.87 & 3.65 ± 3.16 & 79.94 ± 2.99 & 66.15 ± 6.51 & 14.74 ± 8.06 & 3.50 ± 1.44 \\
        Sacrum & 69.20 ± 4.18 & 57.84 ± 5.03 & 12.29 ± 2.72 & 3.92 ± 0.65 & 71.05 ± 4.97 & 53.55 ± 3.86 & 10.20 ± 2.56 & 2.92 ± 0.21 \\
        Vertebrae & 77.38 ± 4.85 & 65.87 ± 2.19 & 10.06 ± 3.31 & 2.84 ± 0.96 & 72.10 ± 2.22 & 54.92 ± 0.78 & 11.06 ± 6.49 & 2.40 ± 0.13 \\
        Heart & 81.06 ± 4.30 & 72.59 ± 5.35 & 12.11 ± 3.56 & 4.10 ± 2.07 & 82.51 ± 1.57 & 69.92 ± 1.59 & 10.97 ± 0.86 & 3.71 ± 0.46 \\
        Aorta & 76.38 ± 3.18 & 59.63 ± 2.04 & 12.20 ± 2.84 & 4.34 ± 0.75 & 72.32 ± 4.66 & 58.12 ± 2.10 & 10.44 ± 1.85 & 2.91 ± 0.20 \\
        Clavicula & 63.34 ± 5.09 & 54.53 ± 4.38 & 15.78 ± 7.34 & 5.57 ± 2.76 & 59.08 ± 5.49 & 40.95 ± 3.75 & 10.39 ± 3.09 & 2.85 ± 0.10 \\
        Femur & 77.05 ± 2.46 & 66.71 ± 3.94 & 10.88 ± 4.34 & 3.38 ± 1.94 & 88.58 ± 1.67 & 79.91 ± 2.05 & 5.74 ± 1.40 & 2.53 ± 0.95 \\
        Hip & 76.50 ± 4.00 & 65.70 ± 3.78 & 10.54 ± 6.84 & 3.77 ± 1.64 & 76.86 ± 2.08 & 62.75 ± 2.60 & 8.67 ± 2.00 & 2.70 ± 0.19 \\
        Autochthon & 75.92 ± 2.06 & 63.69 ± 2.75 & 8.28 ± 4.34 & 3.16 ± 1.55 & 85.05 ± 3.70 & 74.08 ± 3.22 & 5.56 ± 0.61 & 2.07 ± 0.12 \\
        Kidney & 75.61 ± 4.79 & 64.35 ± 3.97 & 14.35 ± 5.06 & 4.78 ± 1.62 & 75.81 ± 3.62 & 62.07 ± 3.47 & 11.03 ± 0.58 & 3.30 ± 0.28 \\
        \cellcolor{lightgreen}Brain & \cellcolor{lightgreen}94.63 ± 3.65 & \cellcolor{lightgreen}91.12 ± 2.39 & \cellcolor{lightgreen}4.58 ± 1.79 & \cellcolor{lightgreen}1.87 ± 0.68 & \cellcolor{lightgreen}95.64 ± 0.87 & \cellcolor{lightgreen}91.65 ± 0.70 & \cellcolor{lightgreen}4.57 ± 0.23 & \cellcolor{lightgreen}1.86 ± 0.15 \\ 
        Liver & 86.52 ± 2.46 & 78.78 ± 3.57 & 14.82 ± 5.33 & 5.14 ± 2.25 & 85.19 ± 1.97 & 72.30 ± 3.48 & 19.80 ± 5.50 & 5.56 ± 1.38 \\
        Stomach & 63.90 ± 8.16 & 52.76 ± 7.49 & 23.24 ± 12.56 & 7.59 ± 2.01 & 69.76 ± 8.80 & 57.94 ± 8.86 & 29.92 ± 11.75 & 7.50 ± 2.44 \\
        Pancreas & 52.61 ± 2.33 & 40.64 ± 5.94 & 18.80 ± 7.66 & 5.45 ± 2.29 & 54.58 ± 5.50 & 30.58 ± 6.94 & 21.38 ± 1.94 & 4.67 ± 0.42 \\ 
        Lung & 89.33 ± 0.80 & 84.72 ± 2.01 & 12.10 ± 2.48 & 4.21 ± 1.50 & 91.88 ± 1.74 & 84.55 ± 2.03 & 7.81 ± 0.70 & 2.57 ± 0.25 \\
        \cellcolor{lightyellow}Esophagus & \cellcolor{lightyellow}50.78 ± 2.14 & \cellcolor{lightyellow}37.89 ± 6.45 & \cellcolor{lightyellow}19.49 ± 9.56 & \cellcolor{lightyellow}7.32 ± 5.32 & 56.96 ± 2.44 & 41.51 ± 4.13 & 9.00 ± 1.74 & 2.33 ± 0.17 \\ 
        \midrule
        \rowcolor{lightgray}Mean & 73.11 ± 1.02 & 62.65 ± 3.07 & 14.15 ± 2.04 & 4.68 ± 1.01 & 73.97 ± 2.26 & 60.31 ± 2.72 & 14.00 ± 2.17 & 3.72 ± 0.46 \\ 
        \bottomrule
    \end{tabular}
    }
\end{table}

\section{Experiment and Results}
\subsection{Datasets and Implementation Details}
Excluding significant misalignments, we used 52 \textsuperscript{18}F-FDG and 60 \textsuperscript{68}Ga-FAPI tracer scans acquired from Nanfang Hospital Southern Medical University on a UIH uEXPLORER (Total-Body) PET/CT to validate LDOS. The sampling times were set to 300 seconds (100\% dose) for FDPET and 15 seconds (5\% dose) for ultra-LDPET. Each dataset included 18 segmented organs and was split into training (80\%), validation (10\%), and testing (10\%) subsets. Five-fold cross-validation was performed separately per tracer. Notably, direct organ annotation on PET images was challenging due to constraints in the clinical workflow. Consequently, we utilized annotations from corresponding CT scans as the gold standard, which were then manually refined. Once trained, our model performs organ segmentation on PET images alone, without reliance on the CT modality.

LDOS was implemented based on the nnU-Net and trained from scratch. We trained 500 epoches. The patch size was \(192 \times 192 \times 192\), and the batch size was 2. Training was conducted on an NVIDIA GeForce RTX 4090 GPU with 24 GB memory.

\begin{table}[h]
    \centering
    \caption{Performance of the 18 organs segmentation results on real LDPET and FDPET images. Dice (\%) is used as the metric.}
    \label{tab:comparison_results}
    \setlength{\tabcolsep}{8pt} 
    \renewcommand{\arraystretch}{1.2} 
    \resizebox{0.85\textwidth}{!}{ 
        \begin{tabular}{@{\hskip 7pt}l@{\hskip 7pt}c@{\hskip 7pt}c@{\hskip 7pt}c@{\hskip 7pt}c@{\hskip 7pt}c@{\hskip 7pt}c@{\hskip 7pt}c@{}}
            \toprule
            \multirow{2}{*}{\textbf{Metric}} & \multicolumn{3}{l}{\textsuperscript{18}F-FDG} & \multicolumn{3}{l}{\textsuperscript{68}Ga-FAPI} \\ 
            \cmidrule(lr){2-4} \cmidrule(lr){5-7}
            & FDPET & LDPET & LDOS & FDPET & LDPET & LDOS \\ 
            \midrule
            Dice & \textcolor{black!50}{76.03 ± 3.61} & 68.03 ± 2.35 & \textbf{73.11 ± 1.02} & \textcolor{black!50}{74.65 ± 3.09} & 65.73 ± 2.79 & \textbf{73.97 ± 2.26} \\
            IoU & \textcolor{black!50}{62.98 ± 4.01} & 50.77 ± 7.39 & \textbf{62.65 ± 3.07} & \textcolor{black!50}{63.68 ± 2.45} & 48.70 ± 8.31 & \textbf{60.31 ± 2.72} \\
            HD95 & \textcolor{black!50}{15.71 ± 3.50} & 21.05 ± 3.84 & \textbf{14.15 ± 2.04} & \textcolor{black!50}{11.03 ± 2.64} & 23.94 ± 5.33 & \textbf{14.00 ± 2.17} \\
            ASD & \textcolor{black!50}{3.96 ± 1.67} & 6.88 ± 1.78 & \textbf{4.68 ± 1.01} & \textcolor{black!50}{3.12 ± 1.06} & 7.06 ± 2.00 & \textbf{3.72 ± 0.46} \\ 
            \bottomrule
        \end{tabular}
    }
\end{table}

\begin{figure*}[t]
\centering
\includegraphics[width=0.75\textwidth]{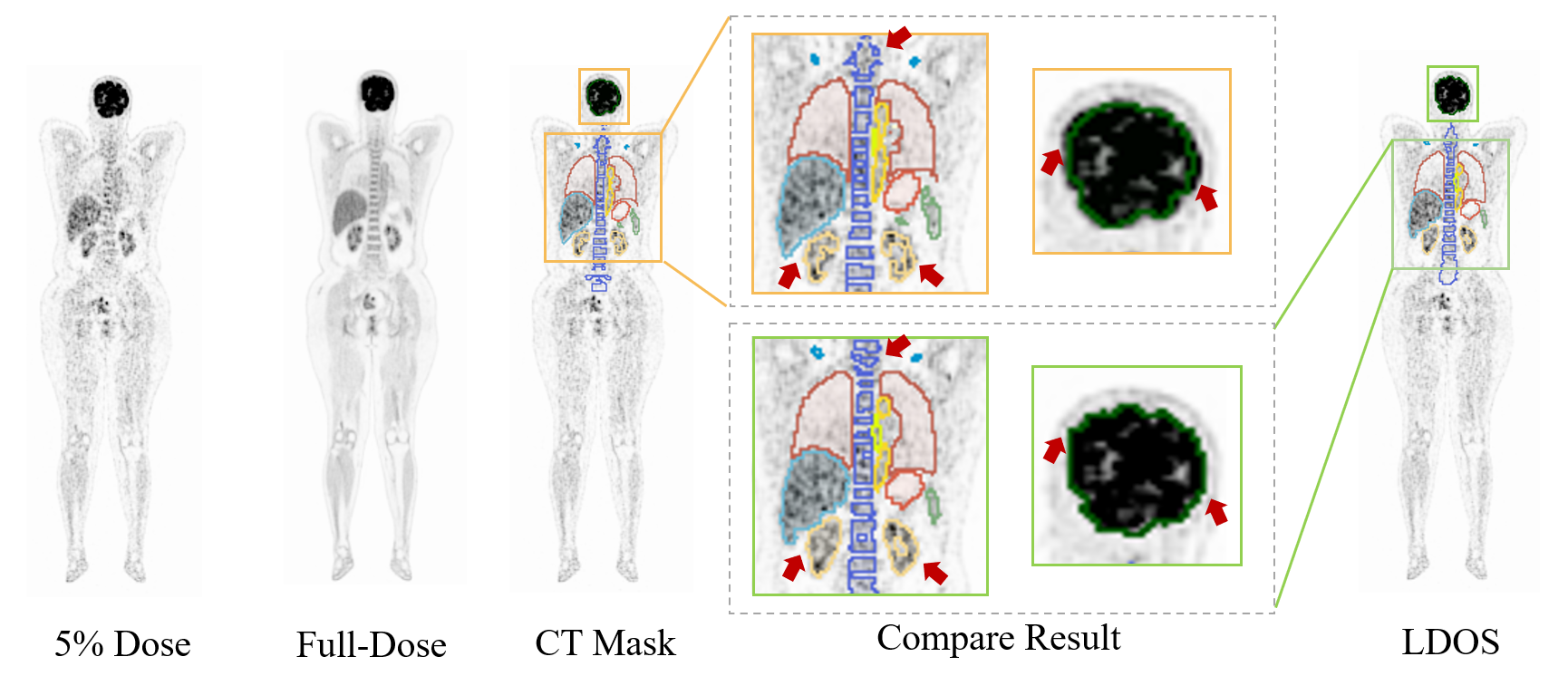} 
\caption{A sample of head and abdominal organs segmentation on the \textsuperscript{18}F-FDG. For clarity, segmentation results for some organs are not displayed.}
\label{fig3}
\vspace{-0.5cm}  
\end{figure*}

\subsection{Multiorgan Segmentation Results}
The segmentation performance of LDOS was evaluated using multiple metrics, including Dice Similarity Coefficient (Dice), Intersection over Union (IoU), 95th percentile Hausdorff distance (HD95), and Average Surface Distance (ASD). A 5-fold cross-validation was conducted on both datasets. The quantitative segmentation results are shown in Table~\ref{tab:merged_results}. The average Dice scores across the 18 organs were \(73.11\% \pm 1.02\%\) and \(73.97\% \pm 2.26\%\) for the \textsuperscript{18}F-FDG and \textsuperscript{68}Ga-FAPI datasets, respectively. Additionally, the average HD95 were \(14.15 \, \text{mm} \pm 2.04 \, \text{mm}\) and \(14.00 \, \text{mm} \pm 2.17 \, \text{mm}\), demonstrating the robust segmentation capability of LDOS at ultra-LPPET (5\% dose).

\begin{figure*}[t]
\centering
\includegraphics[width=1\textwidth]{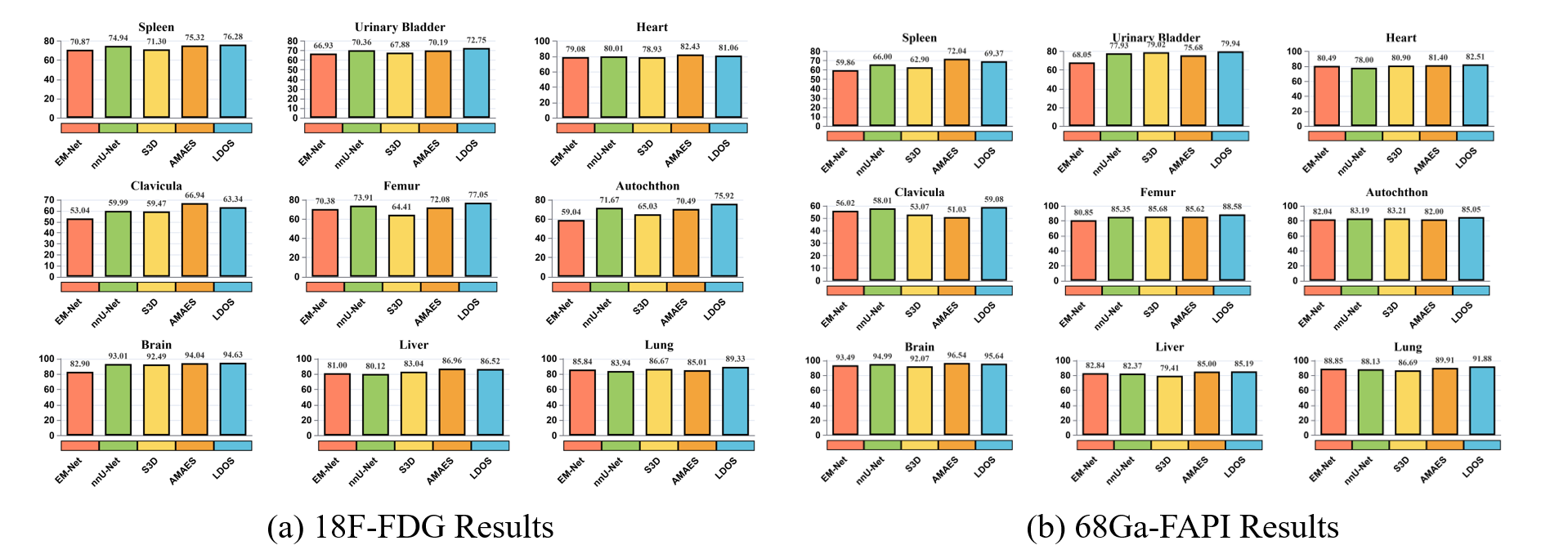} 
\caption{Performance comparison of selected organs across different methods on the \textsuperscript{18}F-FDG (a) and \textsuperscript{68}Ga-FAPI (b) datasets. Dice (\%) is used as the metric.}
\label{fig4}
\end{figure*}

LDOS was compared to the segmentation results of real LDPET and FDPET. For a fair comparison, the same nnU-Net architecture was employed, excluding the denoising decoder loss. As shown in Table~\ref{tab:comparison_results} and Fig.~\ref{fig4}, LDOS achieved notable improvements in Dice scores, with \(5.08\%\) and \(8.24\%\) higher accuracy than LDPET on the \textsuperscript{18}F-FDG and \textsuperscript{68}Ga-FAPI datasets, respectively. Moreover, LDOS achieved segmentation accuracy comparable to or surpassing FDPET for certain organs.

\begin{table}[h]
    \centering
    \caption{Performance comparison of different methods on the \textsuperscript{18}F-FDG and \textsuperscript{68}Ga-FAPI datasets. Dice (\%) is used as the metric. The best performance is highlighted in bold, and the suboptimal performance is underlined.}
    \label{tab:comparison_methods}
    \setlength{\tabcolsep}{8pt} 
    \renewcommand{\arraystretch}{1.2} 
    \resizebox{0.75\textwidth}{!}{ 
        \begin{tabular}{@{\hskip 7pt}l@{\hskip 7pt}l@{\hskip 7pt}c@{\hskip 7pt}c@{}}
            \toprule
            Method & Pre-train Data & \textsuperscript{18}F-FDG & \textsuperscript{68}Ga-FAPI \\ 
            \midrule
            \textcolor{black!50}{Baseline (ResEncU)} & - & \textcolor{black!50}{$68.03 \pm 2.35$} & \textcolor{black!50}{$65.73 \pm 2.79$} \\
            EM-Net~\cite{chang2024net} & - & $65.45 \pm 5.11$ & $62.77 \pm 3.46$ \\ 
            nnU-Net~\cite{isensee2021nnu1} & - & $67.58 \pm 7.30$ & $63.96 \pm 4.15$ \\ 
            & Autopet\uppercase\expandafter{\romannumeral3} & $68.86 \pm 3.61$ & $64.86 \pm 1.98$ \\                                
            \multirow{1}{*}{S3D (nnU-Net)~\cite{wald2024revisiting}} 
                & Our Datasets & $70.35 \pm 1.78$ & $70.87 \pm 2.35$ \\ 
                & Autopet\uppercase\expandafter{\romannumeral3} and Our Datasets & $70.03 \pm 0.96$ & \underline{$72.76 \pm 2.36$} \\
            & - & $67.39 \pm 5.00$ & $64.97 \pm 6.27$ \\ 
            \multirow{1}{*}{AMAES (U-Net B)~\cite{munk2024amaes}}
                & BRAINS-45K & $66.92 \pm 3.76$ & $64.68 \pm 2.99$ \\ 
                & Autopet\uppercase\expandafter{\romannumeral3} & $68.45 \pm 4.37$ & $64.56 \pm 3.89$ \\ 
                & Autopet\uppercase\expandafter{\romannumeral3} and Our Datasets & \underline{$71.01 \pm 1.65$} & $69.33 \pm 1.24$ \\ 
            LDOS & - & \textbf{\boldmath$73.11 \pm 1.02$} & \textbf{\boldmath$73.97 \pm 2.26$} \\ 
            \bottomrule
        \end{tabular}
    }
\end{table}

\begin{figure*}[t]
\centering
\includegraphics[width=0.7\textwidth]{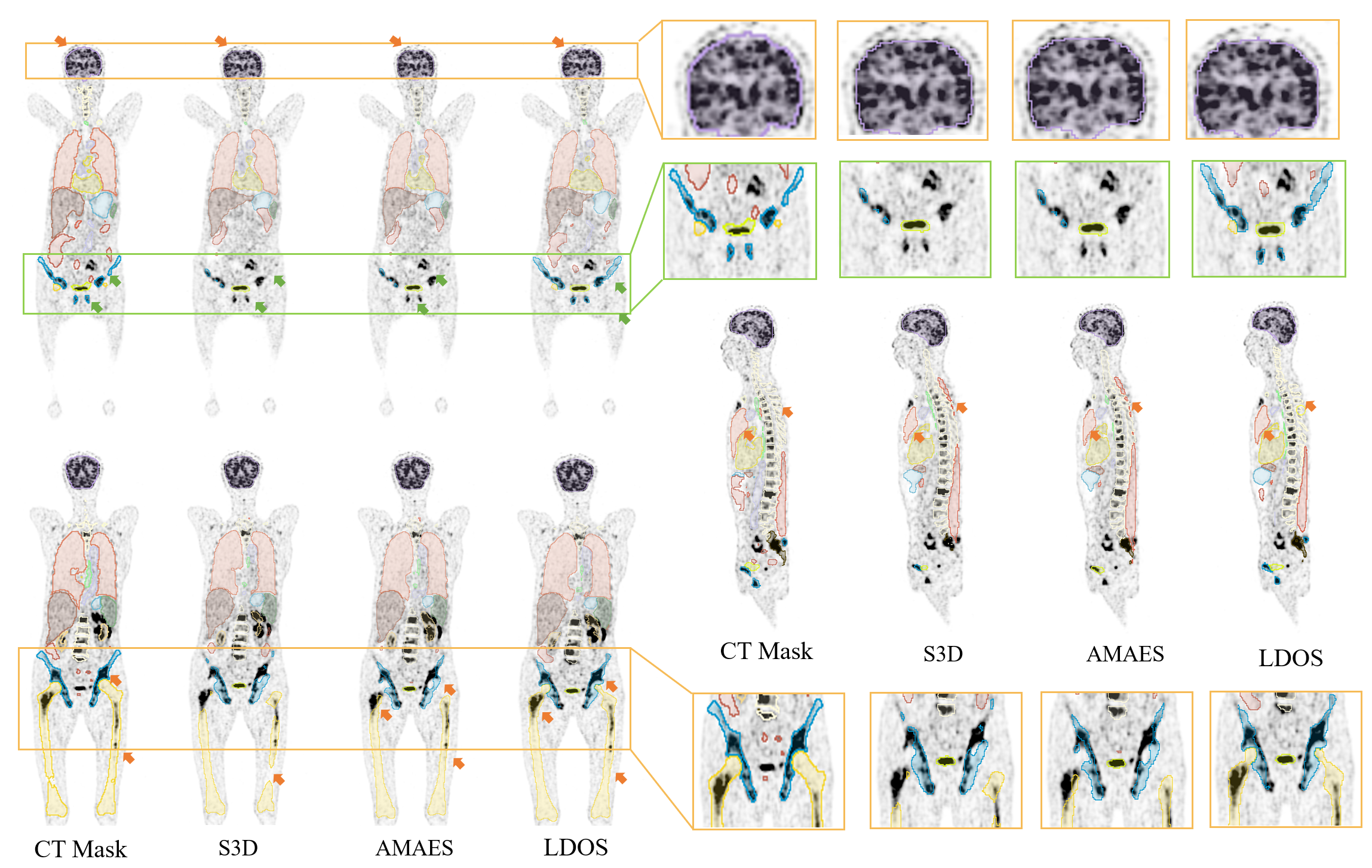} 
\caption{Visualization of segmentation results on the \textsuperscript{18}F-FDG dataset.}
\label{fig5}
\vspace{-0.5cm}  
\end{figure*}

PET organ annotations for model training are derived from corresponding CT images, as direct organ labeling on PET images is impractical. However, achieving precise alignment of training is challenging due to inherent spatial shifts. As illustrated in Fig.~\ref{fig3}, LDOS mitigates the misalignments by semantic feature learning of the self-denoising process, providing a novel solution to improve ultra-LDPET organ segmentation.

\begin{table}[h]
\centering
\caption{Ablation studies of LDOS. LDPET-LDOS/ow means LDOS without the L1 loss weight attenuation strategy, and Fake-FDPET means that denoising and segmentation are performed sequentially. The best performance is highlighted in bold.}
\label{tab:ablation_studies_transposed}
\setlength{\tabcolsep}{4pt} 
\renewcommand{\arraystretch}{1.3} 
\resizebox{0.75\textwidth}{!}{ 
\begin{tabular}{lcccccc}
\toprule
Method       & Dataset        & Dice (\%)                  & IoU (\%)                  & HD95  (mm)                & ASD (mm)                \\ 
\midrule
LDPET                 & \textsuperscript{18}F-FDG & $68.03 \pm 2.35$               & $50.77 \pm 7.39$               & $21.05 \pm 3.84$               & $6.88 \pm 1.78$               \\ 
Fake-FDPET            & - & \textbf{\boldmath$73.75 \pm 3.02$} & $61.46 \pm 2.96$               & \textbf{\boldmath$13.76 \pm 3.05$} & $5.35 \pm 1.84$               \\ 
LDPET-LDOS/ow         & - & $72.01 \pm 4.00$               & $61.01 \pm 2.66$               & $15.02 \pm 2.59$               & $5.72 \pm 1.55$               \\ 
LDPET-LDOS            & - & $73.11 \pm 1.02$               & \textbf{\boldmath$62.65 \pm 3.07$} & $14.15 \pm 2.04$               & \textbf{\boldmath$4.68 \pm 1.01$} \\ 
\midrule
LDPET                 & \textsuperscript{68}Ga-FAPI & $65.73 \pm 2.79$               & $48.70 \pm 8.31$               & $23.94 \pm 5.33$               & $7.06 \pm 2.00$               \\ 
Fake-FDPET            & - & $73.45 \pm 3.01$               & \textbf{\boldmath$61.05 \pm 4.33$} & $16.54 \pm 3.20$ & $4.07 \pm 1.28$               \\ 
LDPET-LDOS/ow         & - & $71.09 \pm 3.71$               & $57.97 \pm 3.15$               & $17.39 \pm 2.69$               & $3.88 \pm 1.09$               \\ 
LDPET-LDOS            & - & \textbf{\boldmath$73.97 \pm 2.26$} & $60.31 \pm 2.72$               & \textbf{\boldmath$14.00 \pm 2.17$} & \textbf{\boldmath$3.72 \pm 0.46$} \\ 
\bottomrule
\end{tabular}
}
\end{table}

\subsection{Comparisons with Previous Results}
As shown in Table~\ref{tab:comparison_methods}, Fig.~\ref{fig4} and \ref{fig5}, LDOS outperformed the state-of-the-art methods. Pre-training methods based on MAE \cite{wald2024revisiting,munk2024amaes}  did not yield significant improvements. LDOS uses semantic features  extracted during the denoising process to achieve superior results. These findings are consistent with prior studies~\cite{wald2024revisiting,munk2024amaes,rokuss2024fdg,li2024well}, which indicate that pre-training on small-scale datasets provide limited advantages for medical imaging tasks.

\subsection{Ablation Experiment}
To evaluate the contribution of the proposed pipeline, we compared its against LDPET and fake FDPET. As shown in Table~\ref{tab:ablation_studies_transposed}, LDOS improved Dice scores on both the \textsuperscript{18}F-FDG and \textsuperscript{68}Ga-FAPI datasets. Unlike fake FDPET, which employed sequential denoising and segmentation, LDOS's collaborative design achieved superior results without complex denoising model training. LDOS reduces deployment costs while enhancing usability. Additionally, results from the LDPET-LDOS/ow (omits the L1 loss weight attenuation strategy) indicates that focusing on denoising semantic information during the early training stages and prioritizing segmentation optimization in the later stages positively impacts the model's overall performance.

\section{Conclusion}
Ultra-LDPET organ segmentation is critical for reliable tracer uptake assessments and
kinetic measurement, but remains largely unexplored. In this study, we propose LDOS, a CT-free ultra-LDPET organ segmentation pipeline. LDOS reinterprets LDPET as a naturally masked version of FDPET and extracts semantic information through a self-denoising process. By incorporating CT-derived organ annotations into the denoising process, LDOS improves anatomical boundary recognition and mitigates PET/CT misalignments, providing a novel solution for LDPET quantification analysis.

\begin{credits}
\sloppy
\subsubsection{\ackname} This research was funded by the National Natural Science Foundation of China (62371221, 12326616 and 62201245), National High end Foreign Experts Recruitment Plan (G2023030025L)and the Science and Technology Program of Guangdong Province 2022A0505050039.

\subsubsection{\discintname}
The authors have no competing interests to declare that are relevant to the content of this article.
\end{credits}

%
%
%
\bibliographystyle{splncs04}
\bibliography{paper}
%

\end{document}